\documentclass[12pt, A4paper]{article}
\usepackage{graphicx}
\usepackage{amsmath}
\usepackage{amsfonts}
\usepackage{amssymb}
\usepackage{cite}
\usepackage[]{authblk}
\usepackage{caption}
\usepackage{subcaption}
\usepackage{geometry}
\begin{document}
\title{{\Large \bf Nonlinear effects in gluon distribution predicted by GLR-MQ evolution equation at next-to-leading order in LHC data}}

\author{ \small M. Lalung\footnote{Corresponding author: mlalung@tezu.ernet.in}, P. Phukan\footnote{email: pragyanp@tezu.ernet.in} and J. K. Sarma\footnote{email: jks@tezu.ernet.in}}

\affil{{\it HEP Laboratory, Department of Physics, Tezpur University}\\
{\it Tezpur, Assam, 784028, India}}
\date{}

\maketitle

\begin{abstract}
In this work we have solved the nonlinear GLR-MQ evolution equation upto next-to-leading order (NLO)  by considering NLO terms of the gluon-gluon splitting functions and running coupling constant $\alpha_s(Q^2)$. Here, we have incorporated a Regge-like behaviour of gluon distribution in order to obtain a solution of the GLR-MQ equation in the range of $5 GeV^2 \leq Q^2 \leq 25 GeV^2$. We have studied the $Q^2$ evolution of the gluon distribution function $G(x,Q^2)$ and its nonlinear effects at small-x. It can be observed from our analysis that the nonlinearities increase with decrease in the correlation radius R of two interacting gluons, as expected. We have compared our result of $G(x,Q^2)$ as $Q^2$ increases and x decreases, for two different values of R, viz. R= 2 $GeV^{-1}$ and 5 $GeV^{-1}$. We have also checked the sensitivity of the Regge intercept $\lambda_G$ on our results. We compare our computed results with those obtained by the global analysis to parton distribution functions (PDFs) by various collaborations where LHC data have been included viz. PDF4LHC15, NNPDF3.0, ABM12 and CT14. Besides we have also shown comparison of our results with HERA PDF data viz. HERAPDF15. 
\end{abstract}
\section{Introduction}
\label{intro}
\par Parton densities in hadrons play significant roles in understanding standard model processes as well as in the predictions for such processes at accelerators. The particle physics community have been so far successfully able to determine the global fits for parton distribution functions from various experiments in the Relativistic Heavy Ion Collider (RHIC) \cite{r3}, CERN's Large Hadron Collider (LHC) \cite{r4,r5,r6}, ZEUS \cite{r8} at HERA. One of the important and interesting aspects of Quantum Chromodynamics (QCD) is to determine the parton density at very small momentum fraction x, where x is also known as the Bjorken variable. Some of the most interesting phenomena in this small-x region are the increase of the parton density as x approaches to zero, the growth of the mean transverse momentum of parton inside the parton cascade at small-x, and the saturation phenomena \cite{r9,r10} of the parton density. In the small-x region study of the gluon density is particularly important because here gluons are expected to dominate the hadron structure function. Study of the gluon distribution function is very important also because of the fact that it is the basic ingredient in the calculation of various high energy hadronic processes like minijet production, growth of total hadronic processes etc. Even in the high energetic processes like p-p, p-A and A-A collision at RHIC and at CERN's LHC, reliable predictions of these processes depend on the precise knowledge of the gluon distribution at small-x. Knowledge of gluon distribution is also important for the computation of inclusive cross-section of hard processes, collinearly factorizable processes for computing cross-sections of proton-proton scattering etc. Also, in the context of studying about the nuclear PDFs (nPDFs), a reliable knowledge on free proton PDF should be known\cite{n01,n02}. A beautiful discussion on the importance of gluon distribution function and various related phenomenon at high energies is given in Refs. \cite{n03,n05}.
\par So far there have been indirect approaches in determining the gluon distribution by probing electron into proton and thus by measuring the proton structure function $F_2 (x, Q^2)$, where $Q^2$ is the four momentum transferred squared or virtuality of the exchanged virtual photon, H1 \cite{r11} and ZEUS \cite{r8} at HERA made us being able to extract information about the gluon distributions in the formerly unexplored region of x and $Q^2$. This method is however, indirect because at small-x $F_2 (x,Q^2)$ actually probes the sea quark distribution, which are related via the coupled QCD evolution equations to the gluon distribution. Moreover, direct determinations of the gluon distribution can be obtained by reconstructing the kinematics of the hadronic final state in the gluon induced processes. Direct method of determination of gluon density have been carried out using events with $J/\psi$ mesons in the final state \cite{r14} and dijet events \cite{r15}.
\par In the perturbative QCD, the high-$Q^2$ behaviour of parton densities have been successfully studied in the context of deep inelastic scattering (DIS) using the linear Dokshitzer-Gribov-Lipatov-Altarelli-Parisi (DGLAP) evolution equations \cite{r16,r17,r18}. The number density of the partons can be evaluated at large-$Q^2$ by solving the linear DGLAP equations in order to calculate the emission of additional quarks and gluons compared to some initial distribution. The results are then adjusted to fit the experimental data (mainly at small-x) available for the proton structure function $F_2 (x,Q^2)$ measured in DIS, over a large range of values of x and $Q^2$ by adjusting the parameters in the initial distributions. With much phenomenological success approximate solutions of the linear DGLAP evolution equations have been reported in recent years \cite{r22,r23,r24,r26,r27,r27a,r27b,r27c,r27d,r27e}.
\par DIS of electron into proton in particular at the HERA facility in DESY, have provided information about the parton distribution at small-x. It is evident from the data that there is a sharp growth of gluon density as x grows smaller. This phenomenon is well backed up by the solutions of linear DGLAP equation which also predicts sharp growth of gluon density towards small-x. However, the gluon density cannot grow forever in order to not violate the unitarity bound \cite{r28,r29} on the physical cross-sections. Hadronic cross-sections comply with the Froissart Bound \cite{r28,r29} which states that the total cross section does not grow faster than the logarithm squared of the energy i.e. $\sigma_{total}=\pi/{m_{\pi} ^2}\ln(s)^2$, where $m_{\pi}$ is the scale of the range of the strong force. A distinguishing effect known as the gluon recombination is believed to be able to unitarize the cross-sections at high energies. Gluon recombination can also provide explanation to any possible  saturation of the gluon distribution function at small-x. The number of gluons at high energy or small-x will be so large that they will spatially overlap with each other. So, the gluon recombination is also as important as the gluon splitting. Gluon recombination effect can also play important role in studying about the nuclear shadowing corrections at small Bjorken x.\cite{n02} In deriving the DGLAP evolution equation the gluon interaction terms were omitted. These interaction terms induce nonlinear corrections to the linear evolution equation. At small-x these corrections of the initial gluons to the evolutionary amplitude should be taken into account. 
\par The DGLAP evolution equations are able to describe the available experimental data quite well in a fairly broad range of x and $Q^2$ with appropriate parametrizations. Despite of the impressive success of DGLAP approach there are some distinct issues that appear in trying to make global fits to the data available from H1 collaboration at HERA\cite{r11}, in the large $Q^2$ region $(Q^2>4 GeV^2)$ as well as in the HERA in the region $(1.5 GeV^2 <Q^2 < 4 GeV^2)$. Moreover, in NLO analysis of MRST2001\cite{r30,r31} when both these regions were included, a good overall fit was found but with a negative gluon distribution. In the CTEQ\cite{r32} collaboration, the large-$Q^2$ region fits global fits are in good agreement with the data while the agreement with small-$Q^2$ region becomes worse. The problem with negative gluon distribution also appear in the NLO set CTEQ6M\cite{r31}. The above mentioned problems are significant because they can be a signal of gluon recombination effect towards smaller values of x and (or) $Q^2$ (but still $Q^2 > \Lambda ^2$, $\Lambda$ being the QCD cutoff parameter) \cite{r33}. These effects thus lead to nonlinear power corrections to the linear DGLAP equations.
\par While considering the gluon interaction terms, the authors of Refs. [31-34] performed detailed study of this region and they have suggested in their pioneering papers\cite{r34,r35,r36} that these shadowing corrections could be well explained within a new evolution equation known as the GLR-MQ equation. The main difference of this equation from the Linear DGLAP equation is the presence of the quantity, $G^2 (x,Q^2)$ which is interpreted as the two-gluon distribution per unit area of hadron. The main features of GLR-MQ equations are that it predicts saturation of the gluon density towards small-x, it predicts a critical line which separates the perturbative regime from the saturation regime and its validity lies in the borders of this critical line \cite{r33,r37}. One of the interesting aspects of the GLR-MQ equation is that it introduces a characteristic momentum scale $Q_s^2$, which measures the density of the saturated gluons. Gribov et al. first suggested a nonlinear evolution equation, they interpreted the evolution kernels as gluon recombination functions at the double-leading logarithmic approximation (DLLA) in a covariant perturbation framework \cite{r38}.
\par GLR-MQ equation is regarded as a `key' link from perturbation region to the non-perturbation region. There has been much work in recent years inspired by GLR-MQ approach that explain the saturation of gluon density at small-x. The `key' ingredient is the gluon recombination in this GLR-MQ approach.
In Ref. \cite{r33} the predictions for the gluon saturation scale using GLR-MQ equation were studied. Some studies of the GLR-MQ equation in the framework of extracting the PDF, of the free proton is given in Refs. \cite{r39,r40,rnew1}.Various studies in the solutions and viable generalisations of the GLR-MQ equation have been reported in recent years \cite{r39,r41}.
\par The solution of GLR-MQ equation is significant for understanding the nonlinear effects which arise due to the gluon-gluon fusion at small-x high gluon density.
In our previous work \cite{r42} (related works Refs. \cite{r44,r45,related,related1}) we obtained a solution of the $Q^2$ and x dependence of gluon distribution from GLR-MQ equation in leading order(LO). In our previous work we addressed the question about the validity of the well known Regge-like parametrizations in the region of moderate-$Q^2$. In this work we will restrict ourselves to moderate-$Q^2$ region and try to obtain a solution of the GLR-MQ equation upto next-to-leading order(NLO) and will try to compare our $Q^2$ evolution result with the results obtained by Global analysis of PDF's by various collaborations like PDF4LHC15\cite{r46}, NNPDF3.0\cite{r6}, HERAPDF1.5\cite{r48}, CT14\cite{r49} and ABM12\cite{r50}.
\section{Theory}
\label{sec:1}
The GLR-MQ equation is a balance equation in which the increase in the number of gluons in a phase space cell $\Delta(1/x)\Delta ln(Q^2)$ gets balanced by the decrease in gluon density through annihilation process. GLR-MQ equation is based upon the following two processes in the parton cascade:
\begin{itemize}
\item The probability of emission of gluons by the vertex G $\rightarrow$ GG ($\propto \alpha_s \rho$) 
\item The probability of induced by annihilation of a gluon by the same QCD vertex GG $\rightarrow$ G ($\propto \alpha_s^2 r^2 \rho^2$)
\end{itemize}
where	 $\rho$($ = xg(x,Q^2)/S_\perp$) is the density of the gluons(with transverse size 1/Q) in the transverse plane and $S_\perp= \pi R^2$. R is the correlation radius between two interacting gluons inside the hadron. The size of the gluon `r' produced during the annihilation process in DIS is proportional to 1/$Q^2$.
In terms of the density of gluons, GLR-MQ equation(with transverse size $1/Q$) in the transverse plane is given by \cite{r34,r35,r36,r51}
\begin{equation}
\frac{\partial^2 \rho}{\partial ln(1/x) \partial ln Q^2} = \frac{\alpha_s N_c}{\pi}\rho-\frac{\alpha_s ^2 \gamma \pi}{Q^2}\rho^2 ,
\end{equation}
 
\par At $x\sim1$ only the production of new gluons(emission) is essential because $\rho<<1$, but at $x\rightarrow0$ the gluon density $\rho$ grows up and the recombination of gluons become significant. 
In terms of the gluon distribution function equation (1) can be expressed as
\begin{equation}
\begin{split}
\frac{\partial^2 xg(x,Q^2)}{\partial ln(1/x) \partial ln Q^2}= \frac{\alpha_s N_c}{\pi}xg(x,Q^2)-\frac{\alpha_s ^2 \gamma}{Q^2 R^2}[xg(x,Q^2)]^2
\end{split}
\end{equation}
\par The value of the factor $\gamma$ was calculated by Mueller and Qiu \cite{r35}. They found $\gamma=\frac{81}{16}$ for $N_c=3$.
Eq. (2) can be expressed in a convenient form using $G(x,Q^2)=xg(x,Q^2)$ which represents the gluon distribution,
\begin{equation}
\begin{split}
\frac{\partial G(x,Q^2)}{\partial ln Q^2} = \frac{\partial G(x,Q^2)}{\partial ln Q^2}\bigg|_{DGLAP}
-\frac{81}{16}\frac{\alpha_s^2 (Q^2)}{R^2 Q^2}\int_{x}^{1} \frac{d\omega}{\omega}G ^2 (\frac{x}{\omega}, Q^2)
\end{split}
\end{equation}
\par The first term in the RHS is the usual linear DGLAP term in the double-leading logarithmic approxmation and the second term is nonlinear in gluon density. The quark-gluon emission diagrams are omitted due to their negligible contribution to the gluon-rich small-x region. The negative sign in front of the nonlinear term is responsible for the gluon recombination. The strong growth generated by the linear term gets tamed by this nonlinear term. So, this term describes the shadowing corrections. The size of the nonlinear term depends on the value of R. For $R=R_h$ shadowing corrections are negligibly small, whereas for $R<<R_h$, shadowing corrections are expected to be large enough, $R_h$ being the radius of the hadron in which the gluons are populated \cite{r51}
\par We introduce a variable $t=\ln(Q^2/\Lambda ^2)$, where $\Lambda$ is the QCD cutoff parameter. Eq. (3) then becomes \cite{r52}
\begin{equation}
\begin{split}
\frac{\partial G(x,t)}{\partial t} =\frac{\partial G(x,t)}{\partial t}\bigg|_{DGLAP} 
-\frac{81}{16} \frac{\alpha_s ^2 (t)}{R^2 \Lambda^2 e^t} \int_{x}^{1} \frac{d\omega}{\omega} G^2 (\frac{x}{\omega},t) ,
\end{split}
\end{equation}
where the first term of Eq. (4) is of the form \cite{r53,r54},
\begin{equation}
\frac{\partial G(x,t)}{\partial t}\bigg|_{DGLAP} =  \int_{x}^{1} P_{gg} (\omega) G(\frac{x}{\omega}, t)d\omega.
\end{equation} 
We have ignored the quark contribution to the gluon rich distribution function.
Considering the terms upto NLO the running coupling constant  $\alpha_s (Q^2)$\cite{r55} of QCD is given by, $$\alpha_s (Q^2) = \frac{4\pi}{\beta_0 \ln(Q^2 /\Lambda^2)}\left(1-b\frac{ \ln\left(\ln(Q^2/\Lambda^2)\right)}{\ln(Q^2/\Lambda^2)}\right)$$ where $b=\frac{\beta_1}{\beta_0 ^2},	\beta_0=11-\frac{2}{3} N_f,	\beta_1=102-\frac{38}{3}N_f$. \\ Now in terms of the variable `t' the running coupling constant $\alpha_s(Q^2)$ takes up the following form 
\begin{equation}
\alpha_s (t)\bigg|_{NLO} = \frac{4\pi}{\beta_0 t} (1-b \frac{\ln t}{t}),
\end{equation} 

\par Here we consider the number of colour charges $N_c =3$ and flavor number $N_f = 4$. 
The splitting function $P_{gg} (\omega)$ can also be expanded as powers of $\alpha_s (t)$, considering upto NLO terms, we may write
\begin{equation}
P_{gg} (\omega) = \frac{\alpha_s (t)}{2\pi} P_{gg} ^{(0)} (\omega) + \left( \frac{\alpha_s (t)}{2\pi}\right)^2 P_{gg} ^{(1)} (\omega),
\end{equation}
where $P_{gg} ^{(0)} (\omega)$ and  $P_{gg} ^{(1)} (\omega)$ are the Altarelli-Parisi splitting kernels at one and two loop corrections respectively as \cite{r17,r17a,r17b}. The expressions of these splitting functions $P_{gg} ^{(0)} (\omega)$ and $P_{gg} ^{(1)} (\omega)$ as given in Refs. \cite{r17a,r17b}  are given below:
\begin{equation*}
\begin{split}
P_{gg} ^{(0)} (\omega ) = 6\bigg( \frac{1-\omega}{\omega} + \frac{\omega}{(1-\omega)_+} + \omega (1-\omega)\bigg)  +(\frac{11}{2}-\frac{N_f}{3})\delta (1-\omega).
\end{split}
\end{equation*} 
Here the denominator of the second term  in the RHS of the above equation is written in terms of '+ prescription' which indicates the cancellation of the divergence that is appearing at $\omega$=1 through 
\begin{equation*}
\int_{0}^{1} d\omega \frac{f(\omega)}{(1-\omega)_+} = \int_{0}^{1} d\omega \frac{f(\omega)-f(1)}{(1-\omega)}.
\end{equation*}
\par where $f(\omega)$ is any arbitrary function. \\
Also,
\begin{equation*}
\begin{split}
P_{gg} ^{(1)} (\omega) = &C_F T_f \bigg\{-16 + 8\omega + \frac{20 \omega ^2}{3} + \frac{4}{3\omega} - (6+10\omega)\ln\omega - 2(1+\omega) \ln^2 \omega\bigg\} \\
&+N_c T_f \bigg\{2 - 2\omega + \frac{26}{9} (\omega ^2 - \frac{1}{\omega})- \frac{4}{3} (1+\omega)\ln\omega - \frac{20}{9} p(\omega) \bigg\} \\&+N_c ^2 \bigg\{ \frac{27}{2} (1-\omega) + \frac{67}{9} (\omega ^2 - \frac{1}{\omega})- ( \frac{25}{3}- \frac{11 \omega}{3} + \frac{44 \omega ^2}{3}) \ln\omega+4(1+\omega)\ln^2 \omega\\ &+(\frac{67}{9} + \ln^2 \omega- \frac{\pi ^2}{3})p(\omega)-4 \ln\omega \ln(1-\omega)p(\omega)+2 p(-\omega) S_2 (\omega)\bigg\}, \\
 p(\omega) =&\frac{1}{1-\omega} + \frac{1}{\omega} - 2 + \omega (1-\omega), \\
 S_2 (\omega) =&\int_{\frac{\omega}{1+\omega}}^{\frac{1}{1+\omega}} \frac{dz}{z} ln(\frac{1-z}{z}) \xrightarrow[\omega]{small} \frac{1}{2}\ln^2 \omega-\frac{\pi ^2}{6}+O(\omega),
\end{split}
\end{equation*}
where the color factor $C_F = (N_c^2 -1)/{2 N_c}$ and $T_f = \frac{1}{2} N_f.$ \\

The DGLAP equation, first term of the RHS of Eq. (4) takes up the following form upto NLO

\begin{equation}
\begin{split}
\frac{\partial G(x,t)}{\partial t}\bigg|_{DGLAP-NLO} = &\frac{3\alpha_{s}(t)}{\pi}\bigg[\bigg\{ \frac{11}{12}-\frac{N_f}{18}+\ln(1-x) \bigg\} G(x,t) \\
 &  +\int_{x}^{1} d\omega \cdot\bigg\{\frac{\omega G(\frac{x}{\omega},t)-G(x,t)}{1-\omega}\bigg\}   \bigg] \\ &+\frac{3\alpha_{s}(t)}{\pi}\int_{x}^{1} d\omega \bigg\{\omega (1-\omega )+\frac{1-\omega}{\omega}\bigg\} G(\frac{x}{\omega},t) \\&+\left( \frac{\alpha_{s}(t)}{2\pi} \right)^2 {I_2}^g (x,t),
\end{split}  
\end{equation}
where 
\begin{equation}{I_2}^g (x,t)= \int_{x}^{1} d\omega \left[ {P^{(1)} _{gg}} (\omega) G(\frac{x}{\omega},t) \right].
\end{equation}
 Now, to simplify our calculations let us guess for a parameter $T_0$ such that 
 $$\left( \frac{\alpha_{s}(t)}{2\pi} \right)^2 = T_0\cdot\frac{\alpha_{s}(t)}{2\pi}.$$  
\begin{equation}
\Rightarrow T^2 (t) = T_0 \cdot T(t)
\end{equation} \\
where $T(t)= \frac{\alpha_s (t)}{2\pi}.$\\

The variation of $T^2(t)$ and $T_0 \cdot T(t)$ with respect to $Q^2$ are compared in Fig. 1 for the range of $5 \leq Q^2 \leq 25$, our range of consideration.


In this work we are considering the moderate range of $Q^2$, where at small-x, the behaviour of parton distribution functions can be well explained in terms of Regge-like behaviour \cite{r59}.
The Regge like behaviour of the sea-quark and anti-quark distribution at small-x is given by $q_{sea}\sim x^{-\alpha_p}$  which corresponds to a pomeron exchange with an intercept of $\alpha_p =1$, whereas the valence quark distribution at small-x is given by $q_{val} \sim x^{-\alpha_R}$ corresponding to reggeon exchange with an intercept of $\alpha_R =0.5$. At moderate $Q^2$, the leading order calculations in $\ln(1/x)$ with fixed value of $\alpha_s$ can be explained in terms of steep power law behaviour of $xg(x,Q^2)$ $\sim x^{-\lambda_G}$, where $\lambda_G = (4\alpha_s N_c/ \pi) \ln2\approx 0.5$ for $\alpha_s = 0.2$, as appropriate for $Q^2 \geq 4 GeV^2$. \cite{regge1,regge2,regge3} Some useful discussions on Regge theory and its applications are given in Ref. \cite{appregge}.
\par Moreover the Regge theory provides extremely naive and simple parameterization of all the total cross sections \cite{cross1,cross2}. Thus it is convenient to use Regge theory as given in Refs. \cite{a,b} for the study of DGLAP evolution equations. We follow the same tactics in order to determine the gluon distribution function with the nonlinear correction using Regge-like behavior \cite{r45} of gluons. The Regge behavior is believed to be valid at small-x and at some intermediate $Q^2$
, where $Q^2$ must be small, but not so small that $\alpha_s(Q^2)$ is too large\cite{w1,w2}. In order to solve the GLR-MQ equation, we will consider a simple form of Regge-like behaviour determining the behaviour of the gluon distribution function at small-x.

We write
\begin{equation}
G(x,t) = \wp (t) x^{-\lambda_G},
\end{equation}
which implies
\begin{equation}
G(\frac{x}{\omega},t) = \wp (t) x^{-\lambda_G} \omega^{\lambda_G}= G(x,t) \omega^{\lambda_G}
\end{equation}
and
\begin{equation}
G^2 (\frac{x}{\omega},t) = (\wp(t) x^{-\lambda_G})^2\omega^{2\lambda_G}= G^2 (x,t) \omega ^{2\lambda_G},
\end{equation}
where $\wp (t)$ is a function of t and $\lambda_G$ is the Regge intercept for gluon distribution function. This form is well supported by the authors in Refs. \cite{r45,w1,w2}. According to Regge theory, small-x behaviour of gluons and sea quarks are controlled by the same singularity factor in the complex plane of angular momentum\cite{r59}. For all the spin independent singlet, non-singlet and gluon distribution functions, the values of Regge intercepts should be close to 0.5 in quite a broad range of small-x as suggested in Ref. \cite{x}, We would also expect that for this value of $\lambda_G= 0.5$, our theoretical results are best fitted to those of experimental data and parametrizations.
Finally the GLR-MQ equation upto NLO becomes
\begin{equation}
\begin{split}
\frac{t}{\left[1-b \ln t/{t}\right]}\cdot \frac{\partial G(x,t)}{\partial t}
=\left[ f(x) + \frac{T_0}{2}\cdot A_f k(x) \right]G(x,t) - T_0\cdot g(x)\cdot  \frac{G^2 (x,t)}{e^t},
\end{split}
\end{equation}
 where $$k(x)=\int_x^1 d\omega P_{gg}^{(1)}(\omega) \omega ^{\lambda_G}$$
and 
\begin{equation*}
\begin{split}
g(x)=&\frac{81}{16} A_f \frac{2 \pi ^2}{R^2 \Lambda ^2} \bigg(\frac{1-x^{2\lambda_G}}{2 \lambda_G}\bigg), \\
f(x)=&3A_f\bigg(\frac{11}{12}-\frac{N_f}{18}+\ln(1-x)+\frac{2}{2+\lambda_G}\bigg)  -3A_f\bigg(\frac{2 x^{\lambda_G+2}}{\lambda_G+2}+\frac{x^{\lambda_G}}{\lambda_G}-\frac{1}{\lambda_G}-x+1\bigg), \\
A_f =&\frac{4}{\beta_0}.
\end{split}
\end{equation*}
Eq. (14) is a partial differential equation, the solution of which is of the form

\begin{equation}
G(x,t)= \displaystyle\frac{e^{\frac{b f_1 (x)}{t}}\cdot t^{(1+\frac{b}{t})f_1 (x)}}{C + \int_{1}^{t} \frac{e^{\zeta (x,z)}\cdot g_1 (x) (z-b \ln z)dz}{z^2}} ,
\end{equation}
where
\begin{equation}
\zeta(x,z)=\frac{b f_1 (x)}{z}-z+f_1 (x) \ln z+b f_1 (x)\frac{\ln z}{z},
\end{equation}
\begin{equation*}
\begin{split}
f_1 (x) =&f(x) +\frac{T_0 A_f k(x)}{2} \\ 
g_1 (x) =& T_0 \cdot g(x) \\
\end{split}
\end{equation*}
where C is a constant to be determined using initial conditions of the gluon distributions for a given $t_0$ or $\ln{(Q_0^2/\Lambda^2)}$, for $Q^2 \geq Q_0 ^2$. Although the Regge like behaviour is not in agreement with the double-leading-logarithmic solution namely $G(x,t)\varpropto e^{[k \ln(t) \ln(1/x)]^{1/2}}$, where k is a constant, the range of small-x and intermediate $Q^2$ is actually the Regge regime, We expect our solution of the form suggested in Eq. (15) is correct in the vicinity of saturation scale, where all our assumptions look natural.
Now, to determine t (or $Q^2$) evolution of  $G(x,t)$ (or $G(x,Q^2)$), we can apply initial conditions at $t=t_0$ for any lower value of $Q^2=Q_0^2$, to get
\begin{equation*}
\begin{split}
G(x,t_0)=\frac{e^{\frac{b f_1 (x)}{t_0}}\cdot t_0^{(1+\frac{b}{t_0})f_1 (x)}}{C+\int_{1}^{t_0} \frac{e^{\zeta (x,z)}\cdot g_1 (x) (z-b \ln z)dz}{z^2}} 
\end{split}
\end{equation*}
and
\begin{equation}
\begin{split}
C=\frac{e^{\frac{b f_1 (x)}{t_0}}\cdot t_0^{(1+\frac{b}{t_0})f_1 (x)}-\int_{1}^{t_0} \frac{e^{\zeta (x,z)}\cdot g_1 (x) (z-b \ln z)dz}{z^2}}{G(x,t_0)} 
\end{split}
\end{equation}
C can be evaluated using Eq. (17) and considering appropriate input distribution $G(x,t_0)$ for a given value of $Q_0^2$. Substituting the expression of C into Eq. (15), we can obtain the t (or $Q^2$) evolution of the gluon distribution function for a fixed-x upto NLO as
\begin{equation}
\begin{split}
G(x,t)= \frac{G(x,t_0)\cdot e^{\frac{b f_1 (x)}{t}}\cdot t^{(1+\frac{b}{t})f_1 (x)}}{{t_0}^{(1+\frac{b}{t_0})f_1 (x)}\cdot e^{\frac{b f_1 (x)}{t_0}}+G(x,t_0)\cdot \int_{t_0}^{t} \frac{e^{\zeta (x,z)}\cdot g_1 (x) (z-b \ln z) dz}{z^2}} \\
\end{split}
\end{equation}
We have thus obtained an expression for the $Q^2$ or t evolution of the gluon distribution function $G(x,t)$ upto NLO by solving the GLR-MQ evolution equation semi-numerically.
\section{Results and Discussions}
In this work we have solved the nonlinear Gribov-Levin-Ryskin-Mueller-Qiu (GLR-MQ) evolution equation semi-numerically upto NLO using Regge ansatz in order to determine the $Q^2$ evolution of the gluon distribution function G(x, $Q^2$). We have shown comparison of our results of $Q^2$ evolution with global fits obtained by various collaborations like NNPDF3.0 \cite{r6} , HERAPDF1.5 \cite{r48} , CT14 \cite{r49} , ABM12 \cite{r50} and PDF4LHC15 \cite{r46}. We have compared our results with the PDF sets in which recent LHC data are included. The NNPDF3.0 set of PDFs uses a global dataset including various HERA data viz. HERA-II deep-inelasic inclusive cross-sections and the combined HERA charm data. Further, they have also included relevant LHC data in their analysis viz. the jet production data from ATLAS and CMS, vector boson rapidity and transverse momentum distributions from ATLAS, CMS and LHCb, W+c data from CMS and the top quark pair production total cross sections from ATLAS and CMS. The QCD fit analysis of the combined HERA-I inclusive deep inelastic cross-sections have been extended to include combined HERA II measurement at high $Q^2$ resulting into HERAPDF1.5 sets. The CT14 PDFs differ from the previous CT PDFs in several aspects, which includes the use of data from LHC experiments, and the new D$\varnothing$ charged lepton asymmetry data. The ABM12 PDF set is successor of their previous ABM11 global fits, resulting from global analysis of DIS and hadron collider data including the available LHC data for standard candle processes such as $W^\pm$ and Z-boson and $t\bar{t}$ production. Finally, PDF4LHC15 PDF set is based on the updated recommendation of PDF4LHC group for the usage of sets of PDFs and the assessment of PDF and PDF+$\alpha_s$ uncertainties suitable for applications at the LHC Run II.

 %


Fig. 1 represents a comparison of variation of $(\alpha_s(t)/{2 \pi})^2$ $ = T^2(t) $ and $T_0 \cdot\alpha_s(t)/{2 \pi}$ with respect to $Q^2$. We guess for the parameter $T_0$ such that the difference between $T^2(t)$ and $T_0 \cdot T(t)$ is minimum. We found $T_0 = 0.036$ for the best fit of the result in the range of $5 GeV^2 \leq Q^2 \leq 25 GeV^2$.
Fig. 2(a-d) represent our best fit results of $Q^2$ evolution of the gluon distribution function G(x, $Q^2$) for R=2 $GeV^{-1}$, computed from Eq. (18) for various values of x viz. $10^{-2},10^{-3},10^{-4}$ and $10^{-5}$ respectively. In all the graphs we have chosen PDF4LHC15 set from which we have picked up the input distribution $G(x,Q_0^2)$ of a given value of initial $Q_0^2$ to predict the $Q^2$ evolution of $G(x,Q^2)$. The input $G(x,Q_0^2)$ is taken at an input value of $Q_0^2$ $\approx$ 5 $GeV^2$. We have taken the input from PDF4LHC15 beacuse this set is based on the LHC experimental simulations, the 2015 recommendations\cite{notun1} of the PDF4LHC working group and contain combinations of more recent CT14,\cite{r49} MMHT2014,\cite{notun2} and NNPDF3.0\cite{r6} PDF ensembles. In this work we have considered the kinematic region to be 5 $GeV^2$ $\leq$ $Q^2$ $\leq$ 25 $GeV^2$ where all our assumptions look natural and our solution seems to be valid. The average value of $\Lambda$ in our phenomenological analysis is taken to be about 0.2 GeV. 
In Fig. 3(a-b), we have investigated the effect of nonlinearity in our results for different values of R and $\lambda_G$. We have compared the gluon distribution function $G(x,Q^2)$ for two different values of R viz. R=2 $GeV^{-1}$ and R=5 $GeV^{-1}$ at various values of x viz. $10^{-2},10^{-3},10^{-4}$ and $10^{-5}$  respectively. The value of R depends on how the gluon ladders are coupled to the proton, or on how the gluons are distributed within the proton. If the gluons are spread throughout the entire nucleon then value of R will be of the order of the proton radius (R $\simeq$ 5 $GeV^{-1}$). On the other hand, if gluons are concentrated in hot-spot then R will be very small (R $\simeq$ 2 $GeV^{-1}$).\cite{naya} We have also checked the sensitivity of the Regge intercept $\lambda_G$ in our result by comparing our result of gluon distribution $G(x,Q^2)$ for three different values of $\lambda_G$ viz. 0.4, 0.5 and 0.6 for x = $10^{-2},10^{-3},10^{-4}$ and $10^{-5}$. 



 %

\section{Conclusion}      
In this work we have incorporated Regge like behaviour of the Gluon distributions in the kinematic range of moderate-$Q^2$ and solved the GLR-MQ evolution equation upto next-to-leading order by considering upto NLO terms of the gluon-gluon splitting function $P_{gg}(\omega)$. We have also incorporated the NLO terms of the running coupling constant $\alpha_s(Q^2)$ into our calculations. From our phenomenological study we can expect that our solution given by Eq. (18) is valid in the kinematic region $5 GeV^2 \leq Q^2 \leq 25 GeV^2$ and $10^{-5} < x < 10^{-2}$, where nonlinear effects cannot be neglected. Like our previous result at LO\cite{r42}, our NLO result of gluon distribution function also increases as $Q^2$ increases and x decreases which is in agreement with perturbative QCD fits at small-x.  It is interesting to see in our phenomenological analysis that the gluon distribution $G(x,Q^2)$ of our NLO solution lies slightly above the LO result as $Q^2$ increases and x decreases. This is due to the contribution from the NLO term in the splitting function $P_{gg}(\omega)$. As we go on decreasing x, the taming of $G(x,Q^2)$ is apparently observed in our NLO solution as seen in Fig. 2(c-d). Through our analysis we have also checked the validity of Regge type behaviour of the gluon distribution function at moderate-$Q^2$ which seems to be valid and compatible with various parametrizations and global fits. We thus can conclude that the solution suggested in Eq. (18) is valid only in the vicinity of the saturation border.

\par It can be observed that our results show almost similar behavior to those obtained from various global parametrizations groups and global fits. However, the nonlinearities play important role for $x\leq10^{-3}$. We have investigated the effect of nonlinearities in our results for different values of R and $\lambda_G$. The gluon distribution function $G(x,Q^2)$  shows steep behaviour at R=5 $GeV^{-1}$ whereas the taming of $G(x,Q^2)$ is more prominent at R=2 $GeV^{-1}$ as $Q^2$ increases and x decreases. Besides, it is also clearly seen that our result is sensitive to the value of Regge intercept $\lambda_G$ as x goes on decreasing. We conclude from our phenomenological analysis that for $x\leq10^{-3}$ our NLO solution is better than the LO result.

\section{Acknowledgements}
One of the authors(M. Lalung) is grateful to CSIR, New Delhi for financial assistantship in the form of Junior Research Fellowship.

\newpage
\begin{figure*}[t]
 \includegraphics[width=0.5\textwidth]{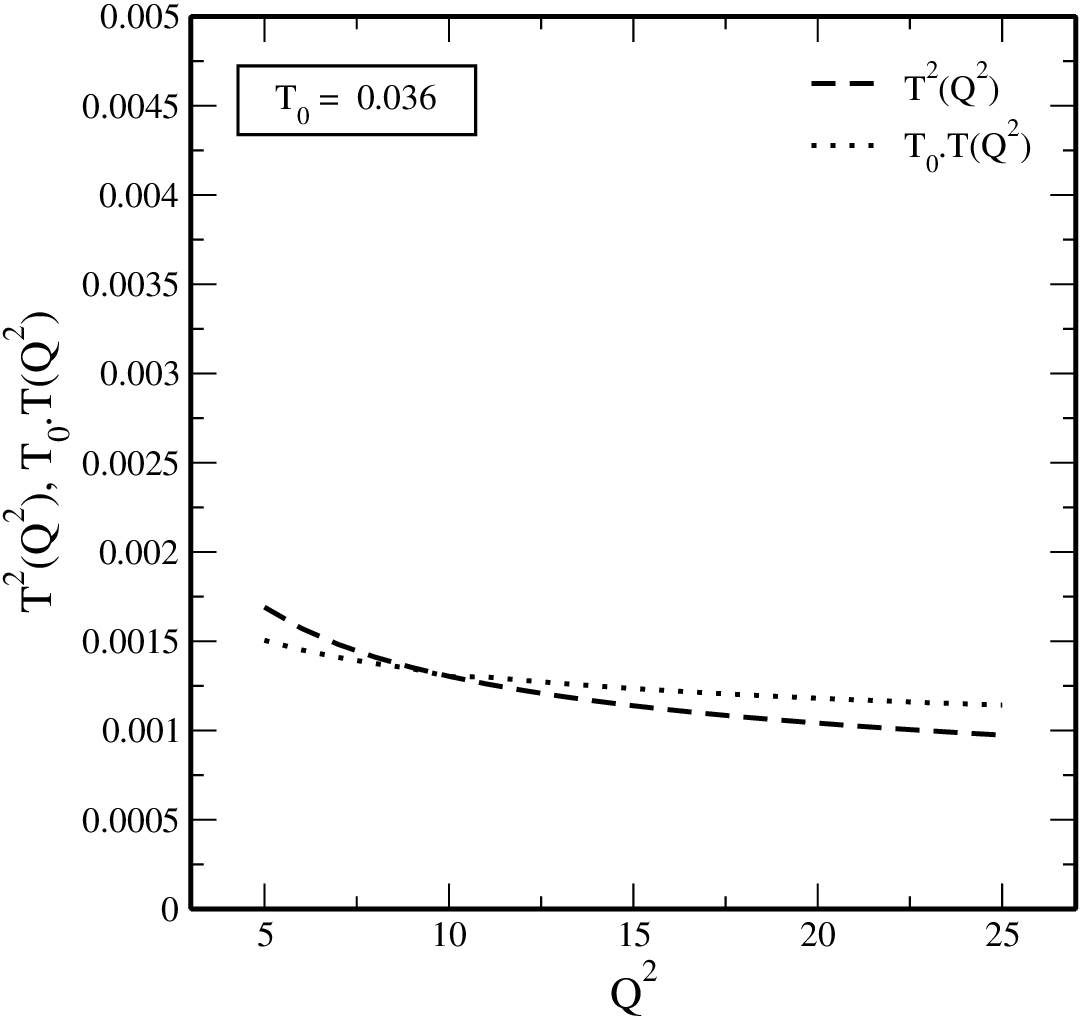}
 \centering
\caption{Variation of $T^2 (Q^2)$ and $T_0\cdot T(Q^2)$ with respect to $Q^2$ in the range $5 GeV^2 \leq Q^2 \leq 25 GeV^2$}
\label{fig:1}       
\end{figure*}

\newpage

\begin{figure}[t] 
  \label{ fig7} 
  \begin{minipage}[b]{0.45\textwidth}
    \centering
    \includegraphics[width=.9\linewidth]{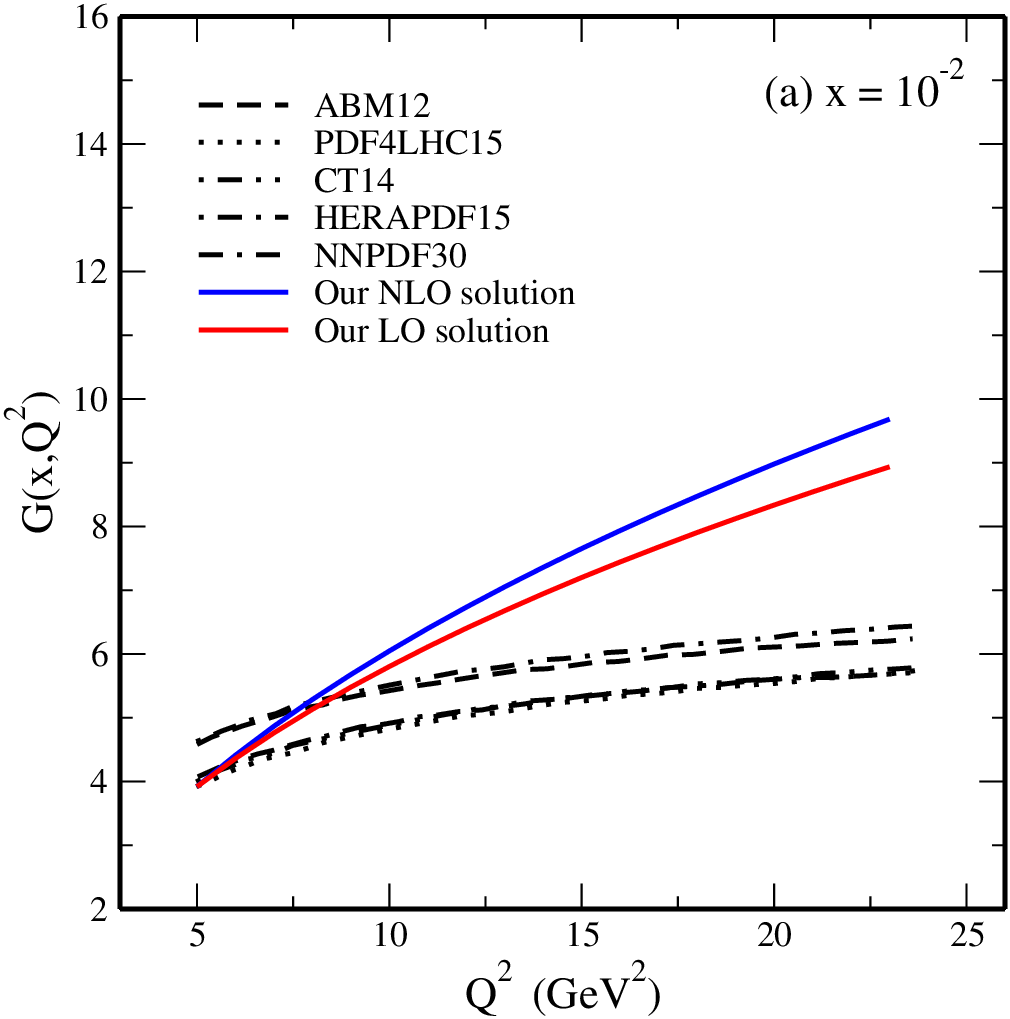} 
    \vspace{8ex}
    
  \end{minipage}
  \hfill
  \begin{minipage}[b]{0.45\textwidth}
    \centering
    \includegraphics[width=.9\linewidth]{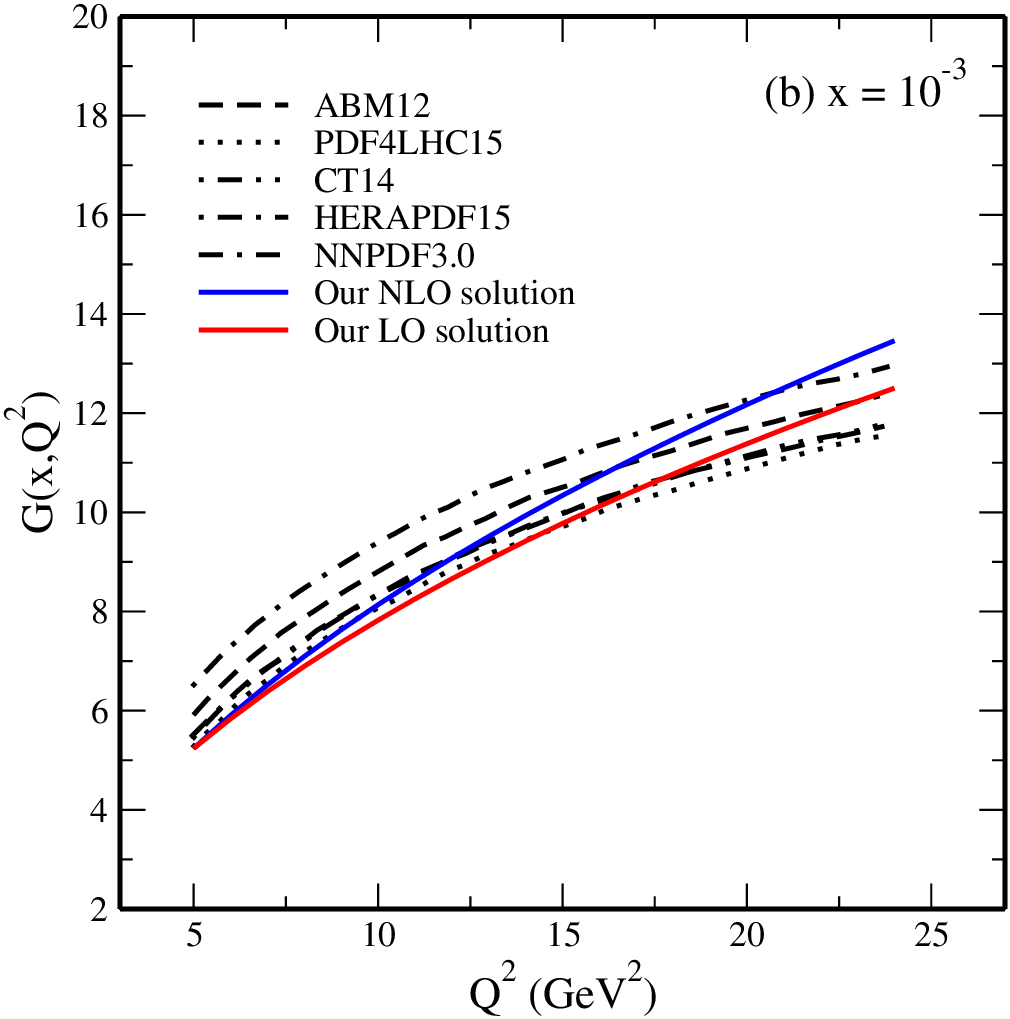} 
    \vspace{8ex}
  \end{minipage}
  \hfill 
  \begin{minipage}[b]{0.45\textwidth}
    \centering
    \includegraphics[width=.9\linewidth]{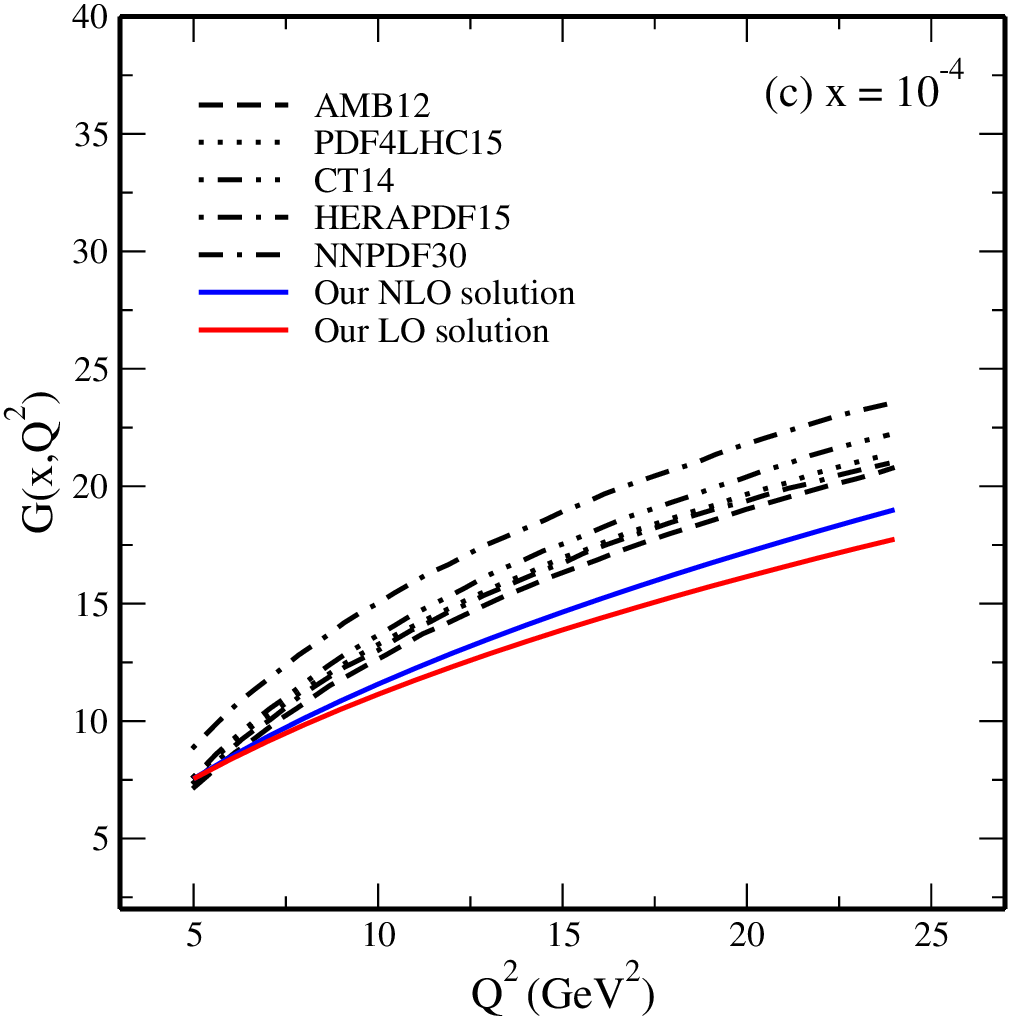} 
    \vspace{8ex}
  \end{minipage}
  \hfill
  \begin{minipage}[b]{0.45\textwidth}
    \centering
    \includegraphics[width=.9\linewidth]{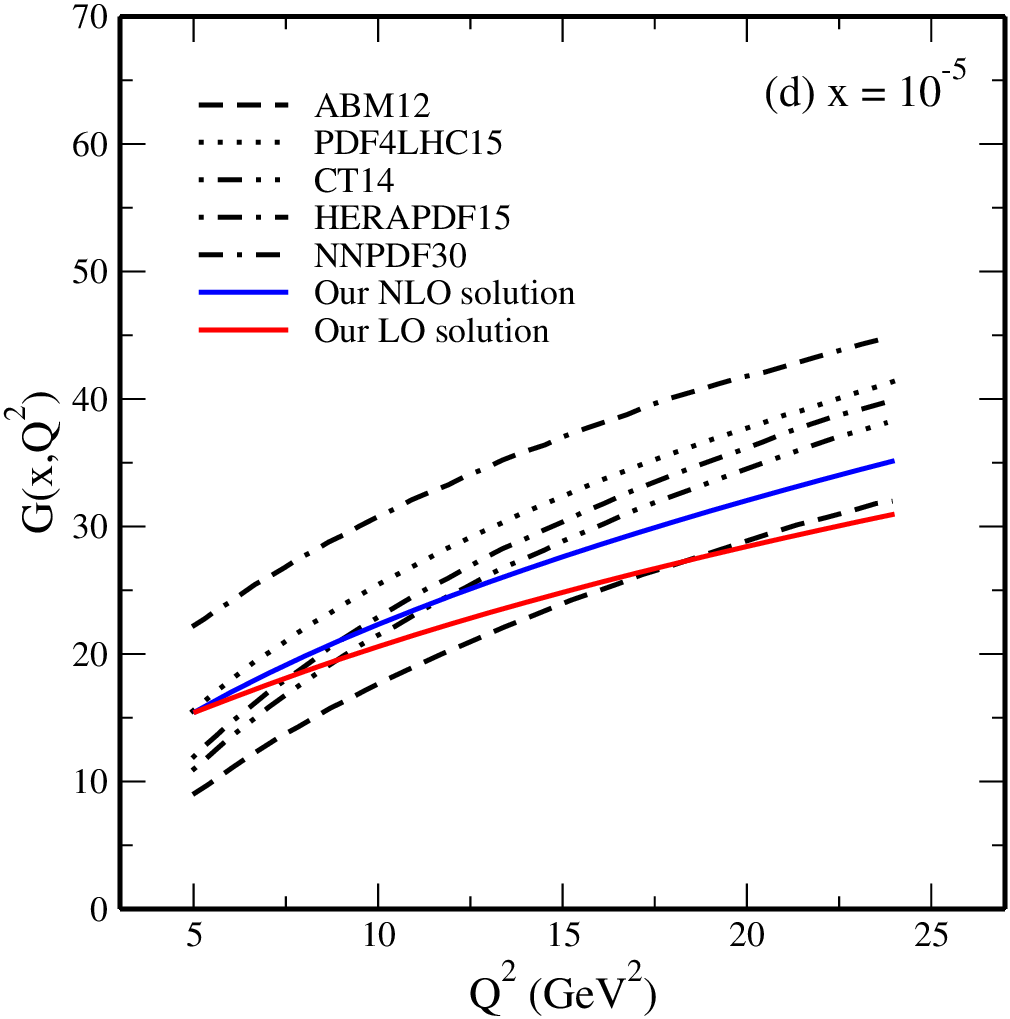} 
    \vspace{8ex}
  \end{minipage} 
  \hfill
  \caption{$Q^2$ evolution of $G(x,Q^2)$ for R= 2 $GeV^{-1}$. \textbf{\emph{Solid blue lines}} are our NLO result while  \textbf{\emph{ solid red lines}} are the LO result,  \textbf{\emph{ dashed lines}} are ABM12 results, \textbf{\emph{dotted lines}} are the result from PDF4LHC15 set,  \textbf{\emph{ dashed dot dot lines}} are from CT14 set,  \textbf{\emph{ dashed dot dashed lines}} are HERAPDF15 set and finally,  \textbf{\emph{dashed dashed dot lines}} are from the NNPDF3.0 set}
\end{figure}

\newpage

\begin{figure}[ht] 
  \label{ fig7} 
  \begin{minipage}[b]{0.45\textwidth}
    \centering
    \includegraphics[width=.9\linewidth]{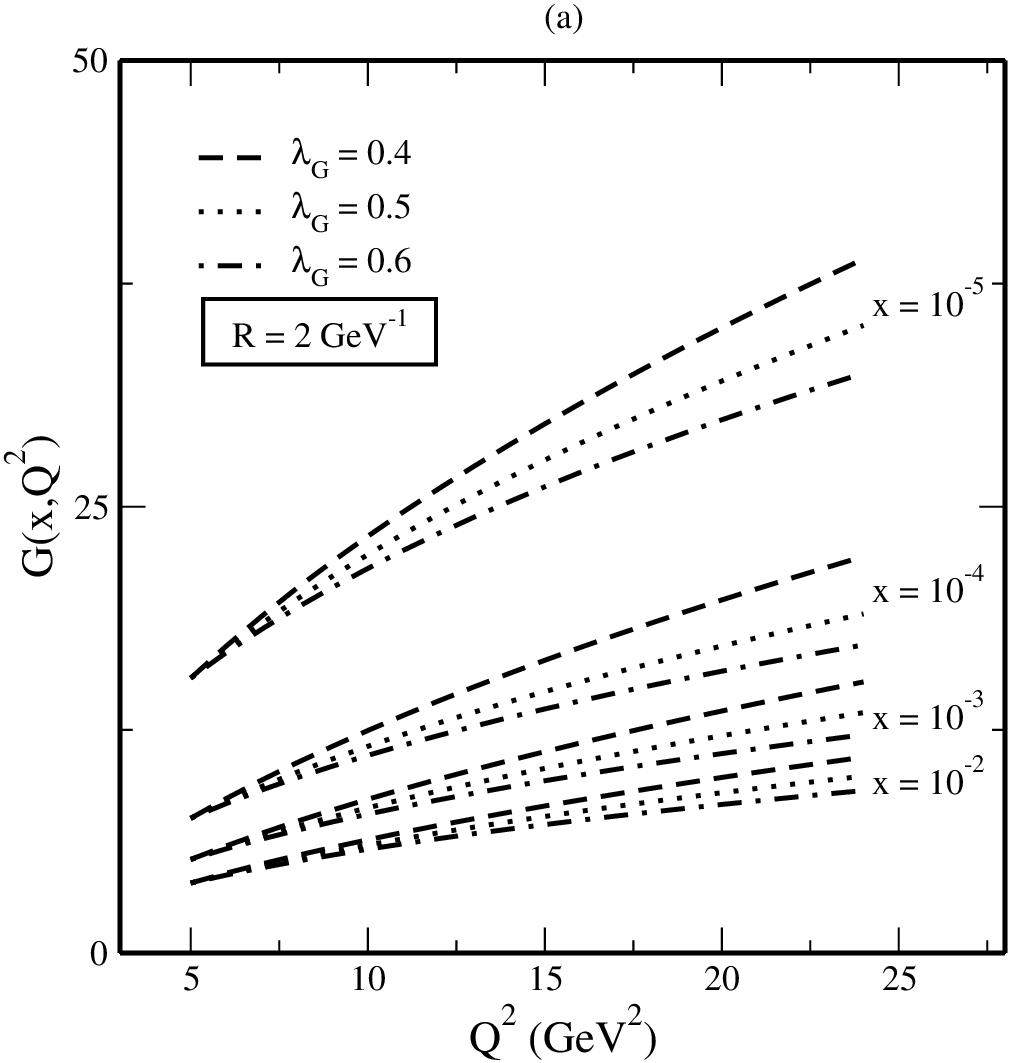} 
    \vspace{8ex}
    
  \end{minipage}
  \hfill
  \begin{minipage}[b]{0.45\textwidth}
    \centering
    \includegraphics[width=.9\linewidth]{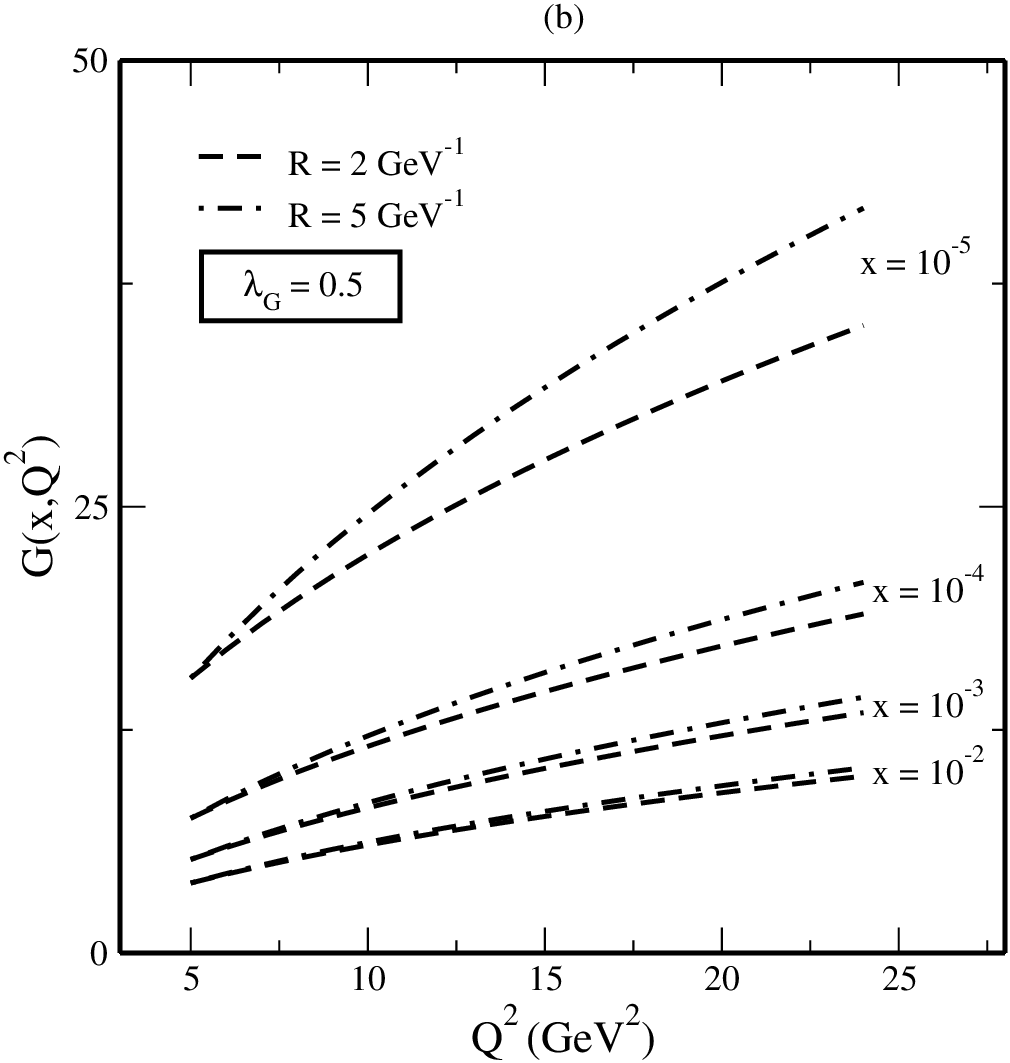} 
    \vspace{8ex}
  \end{minipage}
  \hfill 
   \caption{Sensitivity of our result verses $Q^2$ for different values of R and $\lambda_G$}
\end{figure}

\end{document}